\documentclass[reprint, superscriptaddress, prl, amsmath,amssymb, aps]{revtex4-2}

\pdfoutput=1

\usepackage{graphicx}
\usepackage{dcolumn}
\usepackage{bm}
\usepackage[colorlinks= true, linkcolor=blue, citecolor=blue, urlcolor=blue]{hyperref}
\usepackage{lipsum}
\usepackage{mathtools}
\usepackage{xcolor}
\usepackage{multirow}

\newcommand{\revise}[1]{\textcolor{black}{#1}}

\graphicspath{{./figs/}{./}}

\begin{document}

\title{Stochastic dynamics of granular hopper flows: 
\\a configurational mode controls the stability of clogs}

\author{David Hathcock}
\thanks{D.H. and S.D. contributed equally to this work.} 
\affiliation{IBM T. J. Watson Research Center, Yorktown Heights, NY 10598, USA}
\author{Sam Dillavou}
\thanks{D.H. and S.D. contributed equally to this work.} 
\affiliation{Department of Physics \& Astronomy, University of Pennsylvania, Philadelphia, PA 19104, USA}
\author{Jesse M. Hanlan}
\affiliation{Department of Physics \& Astronomy, University of Pennsylvania, Philadelphia, PA 19104, USA}
\author{Douglas J. Durian}
\affiliation{Department of Physics \& Astronomy, University of Pennsylvania, Philadelphia, PA 19104, USA}
\author{Yuhai Tu}
\affiliation{IBM T. J. Watson Research Center, Yorktown Heights, NY 10598, USA}

\date{\today}

\begin{abstract}

Granular flows in small-outlet hoppers exhibit several characteristic but poorly understood behaviors: temporary clogs (pauses) \revise{where flow stops before later spontaneously restarting}, permanent clogs that last indefinitely, and non-Gaussian, non-monotonic flow-rate statistics. These aspects have been studied independently, but a model of hopper flow that includes all three has not been formulated. Here, we introduce a phenomenological model that provides a unifying dynamical explanation of all three behaviors: coupling between the flow rate and a hidden mode that controls the stability of clogs. In the theory, flow rate evolves according to Langevin dynamics with multiplicative noise and an absorbing state at zero flow, conditional on the hidden mode. The model fully reproduces the statistics of pause and clog events \revise{of} a large ($>40,000$ flows) experimental dataset, including non-exponentially distributed clogging times and non-Gaussian flow rate distribution, and explains the stretched-exponential growth of the average clogging time with outlet size. Further, \revise{we identify the physical nature of the hidden mode in microscopic configurational features}, including size and smoothness of the static arch structure formed during pauses and clogs. Our work provides a unifying framework for several poorly understood clogging phenomena, and suggests numerous new paths toward further understanding of this complex system.
\end{abstract}


\maketitle


Clogging in granular hopper flows is an extreme event, where the formation of a stable arch by a small number of grains near the outlet entirely stops the motion of the bulk material \cite{manna_intermittent_2000, to_jamming_2001,  zuriguel2005jamming, zuriguel2014clogging}. Similar phenomena appear across scales in systems moving through narrow bottlenecks, from suspensions \cite{roussel2007general,marin2018clogging,stoop2018clogging,souzy2020transition,dressaire2017clogging} and collections of cells \cite{dressaire2017clogging,fogelson2015fluid,brescher2013biofilm} to human and animal crowds \cite{helbing2000simulating, helbing2005self, zuriguel2014clogging}. The dependence of flows and clogging on grain properties \cite{ hafez2021effect,gella2022dual, escudero2022kinematics, zuriguel2011silo, zuriguel2005jamming,thomas2013geometry,zuriguel2020statistical} and hopper geometry \cite{zuriguel2020statistical, zuriguel2005jamming, janda2012flow, thomas2013geometry, thomas2015fraction, janda2008jamming, caitano2021characterization,gago2023effect} have been extensively studied. Classical  models~\cite{beverloo_flow_1961,zuriguel_jamming_2003, thomas2015fraction, zuriguel2020statistical} accurately predict many mean behaviors, such as average flow rate as a function of outlet size, but many statistical aspects remain a mystery and clogs cannot be reliably predicted.

As hopper outlet size is decreased, clogs form more quickly and several unexplained features emerge. First, flow rate fluctuations grow (compared to the mean) and become markedly non-Gaussian \cite{janda2009flow, thomas2016intermittency, zhang2023experimental}.
Second, temporary clogs (pauses) emerge, during which a metastable arch forms and flow is temporarily suspended before spontaneously restarting. This behavior has primarily been studied in granular systems perturbed by external vibrations \cite{janda2009unjamming,zuriguel2014clogging, caitano2021characterization,merrigan2018,nicolas2018trap,guerrero2019nonergodicity} or fluid dynamics \cite{souzy2020transition,bertho2003intermittent}, but fully spontaneous unclogging is also possible \cite{janda2009flow, thomas2016intermittency, harth2020intermittent}. 
A single unifying model coupling flow dynamics and intermittent pausing has not yet been proposed.

In this Letter, we introduce a minimal phenomenological model that captures the stochastic dynamics of the flow rate, temporary pauses, and permanent clogs characteristic of granular flow through a small outlet. We find that just two state variables---flow rate and a binary configurational mode---are sufficient to accurately predict the statistics of the flow rate, clogging time, mass ejected, and number of pauses, all of which are confirmed using an experimental dataset of approximately 42,000 flows over five outlet sizes. The flow rate is modeled using Langevin dynamics with multiplicative noise and an absorbing state at zero flow representing pauses and clogs. Our stochastic modeling reveals that the flow rate is coupled to a hidden mode that controls the stability of clogs, resulting in intermittent pauses in the flow, non-Gaussian flow rate distributions and non-exponential clogging times. \revise{This mode represents configurational features; } its dynamics perform independent Bernoulli trials, consistent with previous depictions of flow as randomly sampling positional microstates~\cite{thomas2015fraction}. Our results indicate that tracking a single order-parameter, corresponding to the configurational mode, enables advanced prediction of the long-term stability of an incipient clog.

\begin{figure}[t]
{\centering \includegraphics[width=0.85\linewidth]{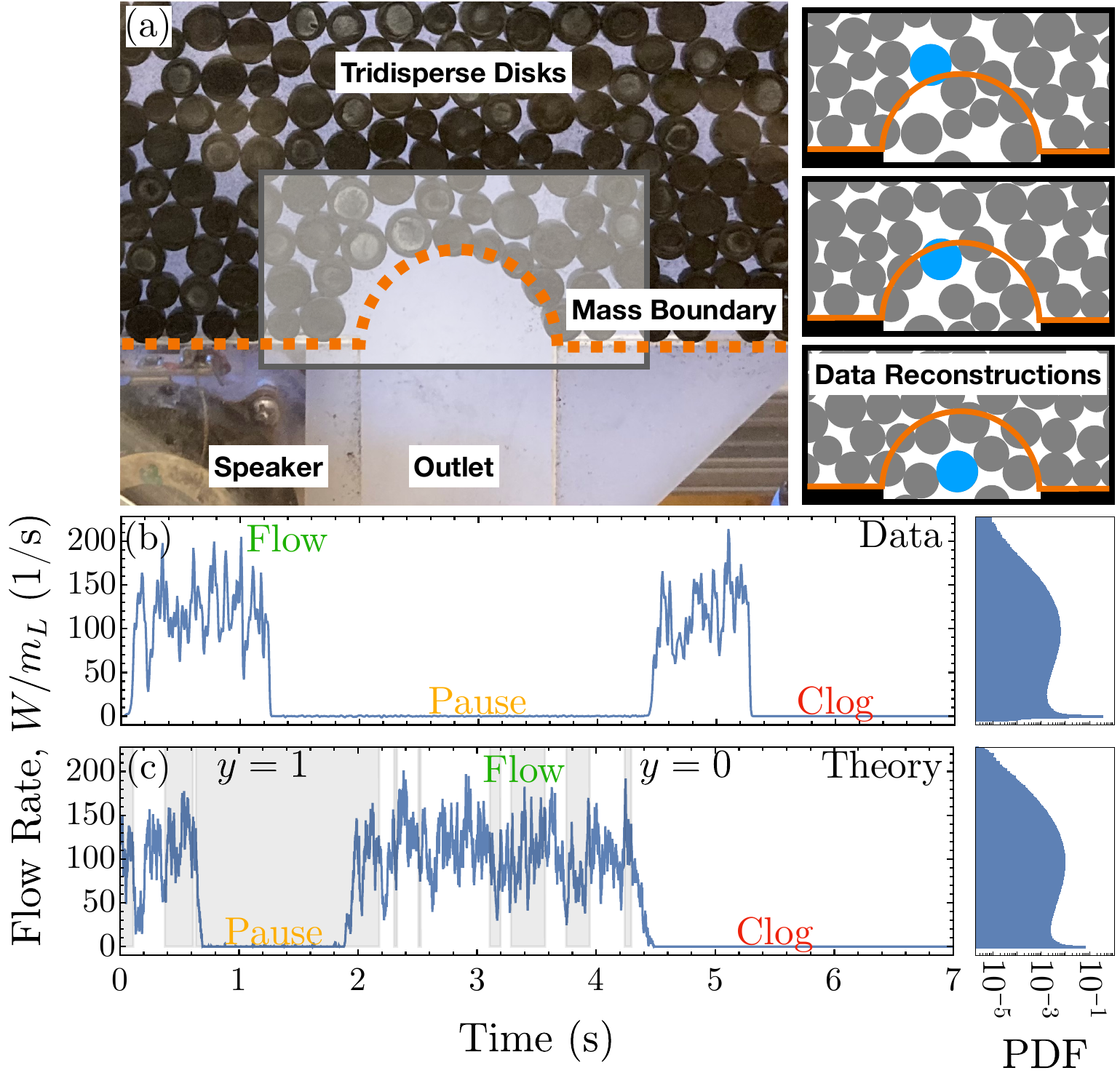}}
\caption{\label{fig:experiment_trajectories_model}(a) Image of the experimental system and snapshots of reconstructed data. The mass boundary line is shown in orange. A large particle is highlighted in blue in the reconstruction, which shows three snapshots each separated by 3 frames ($\Delta t = 3/130 \approx 23$ms). When a grain straddles the mass boundary line, only a fraction of its mass is counted as ejected.  
(b) An experimental flow rate trajectory, normalized by large grain mass, with a single pause. The histogram shows the distribution of flow rates across all flows with outlet size $D= 3.86 d_L$. 
(c) A trajectory and flow rate distribution from simulations of the Langevin model, Eqs.~(\ref{flowRateDynamics}) \& (\ref{hiddenVariable}). The background color indicates the hidden variable state, $y=0$ (white) and $y=1$ (gray).
}
\end{figure}

The experimental data in this study were collected using an automated recirculating hopper. The apparatus, dataset, and procedures are fully described in~\cite{hanlan2024cornerstones}. Briefly, the hopper shown in Fig.~\ref{fig:experiment_trajectories_model}(a) is a vertical quasi-2D plexiglass cell containing an ensemble of tri-disperse anti-static polyethelyne disks with diameters, $d_{S} = 6.0$~mm, $d_{M}= 7.4$~mm, $d_{L} = 8.6$~mm, and equal heights. The disks form a single layer and flow through an opening until they spontaneously clog. Grain positions are recorded by a camera at 130~fps and tracked using custom MATLAB software. Mass ejected over time $M(t)$ is calculated as the total mass of particles that passes a semi-circular boundary spanning and centered on the outlet. This permits ignorance of particles in the center of the outlet, which are difficult to track. Flow rate $W(t)$ is calculated at each frame by taking the slope of a second-order polynomial fit to a five frame window. 

Flows are initiated via mechanical vibration using a speaker pressed against the rear wall of the hopper, near the outlet [bottom left, Fig.~\ref{fig:experiment_trajectories_model}(a)]. The flow is recorded and tracked, and is considered complete when no grains fall through the outlet for five \revise{consecutive} seconds. Ejected grains are directed into a chute to one side, and recirculated into the top of the hopper by a continuous air stream directed upwards. This chute is placed sufficiently far away and shielded so that recirculation air currents do not effect the flow through the outlet. Vents placed along the edges of the hopper allow the injected air to escape. The hose connection from the recirculation motor is, however, a constant source of small vibrations in the hopper, which increases the \revise{likelihood} of unclogging; $38\%$ of flows have at least one pause. This low-level vibration is far smaller than the vibration used to restart the flow. Using this automation scheme, the experiment can continuously run without human intervention. Over 35,000 flows with outlet width $D= 3.86 d_L$ were recorded, along with at least one thousand experiments each for four additional outlet sizes. About $21\%$ of flows pause exactly once before clogging permanently; Fig.~\ref{fig:experiment_trajectories_model}(b) shows a typical trajectory in this category. 

Our goal is to identify a minimal set of macroscopic state variables that determine the collective flow dynamics of the granular material. This approach is similar to modeling gasses with pressure and temperature or identifying a reaction coordinate for chemical reactions~\cite{hanggi1990reaction}. Previous work has shown that granular materials have fluctuations driven by multiplicative noise~\cite{claudin1998models,clark2012particle,clark2014collisional,behringer2014statistical}. In the context of hopper flows, multiplicative noise is also essential for modeling clogs, \textit{i.e.} an absorbing state of the dynamics. 
\revise{Following established studies on} systems with absorbing states~\cite{Grinstein1996Phase,Munoz1997Survival}, 
\revise{we model the flow rate fluctuations }using Langevin dynamics with multiplicative noise. 

\revise{Dynamics for the flow rate alone are not enough to explain pausing and intermittency. Without additional degrees of freedom there are only permenant clogs: it extremely unlikely that the system will spend a long duration at the absorbing state but eventually escape. Thus, modeling the flow rate dynamics reveals the existence of a hidden mode that is coupled to the flow rate but not directly measured by observing a flow-rate trajectory. This hidden variable is a coarse-grained representation of additional grain degrees of freedom beyond the flow rate itself. As we demonstrate later, the hidden mode represents configurational degrees of freedom that determine clog stability: whether a hopper with no flow remains clogged permanently or will restart spontaneously.} 

The instantaneous flow rate normalized by the weight of a large particle, $x= W/m_L$, evolves according to,
\begin{equation}\label{flowRateDynamics}
\dot x = f(x) + \sqrt{2 (g(x) + \epsilon y)} \, \eta(t),
\end{equation}
where $f(x)$ is a deterministic force, $g(x)$ is the flow-rate-dependent noise amplitude, $y$ is the hidden mode state (which increases the noise by $\epsilon$), and $\eta(t)$ is Gaussian white noise~\footnote{Gaussian white noise with mean and correlation $\langle \eta(t) \rangle =0$ and $\langle \eta(t) \eta(t') \rangle = \delta(t-t')$ respectively.}. We require $f(0)=g(0)=0$, so that $x=y =0$ is an absorbing state. We use force $f(x) = -x^2(x-x_0) \Psi(x)/(\tau_0 x_0^2)$, with zeros at the clog $x=0$ and steady flow rate $x_0$~\footnote{We also fit forces of the form $f(x) \propto (x-x_0)(A x + B x^2)$. In all cases the quadratic term was dominant and including the linear term does not considerably reduce the fitting loss.}. Here $\tau_0$ is the correlation time at $x=x_0$ and $\Psi(x)$ is a positive-definite function containing additional nonlinearities \cite{SM}. Expanding to linear order, $g(x) = \Delta \cdot x/x_0$, where $\Delta$ is the noise amplitude. 

\begin{figure}[t!]
\includegraphics[width=0.95\linewidth]{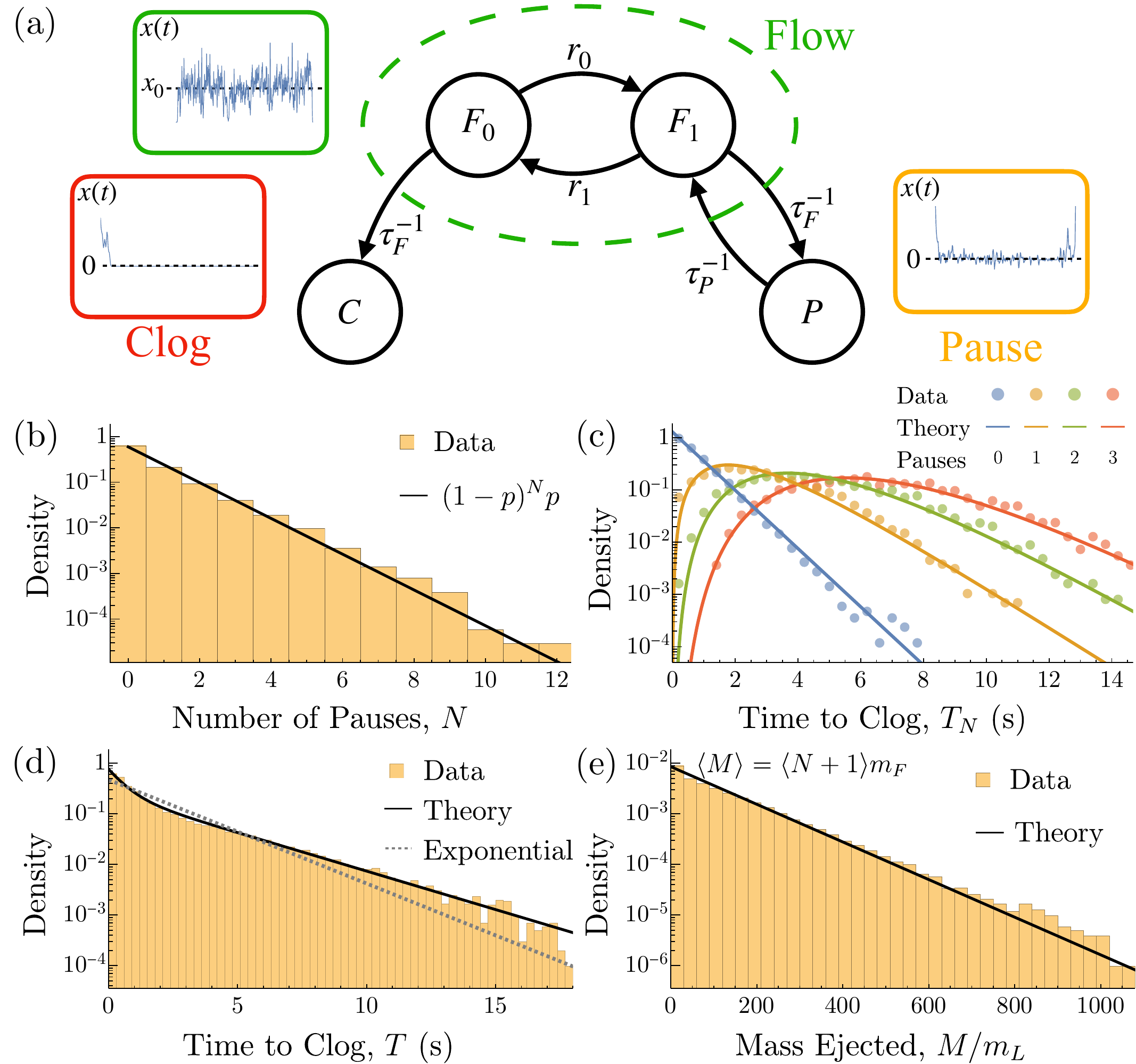}
\caption{\label{fig:flowPauseClogStatistics}
\emph{Predicting flow trajectory statistics.} (a) A four-state representation of our granular flow model. Two flow states $F_0$ and $F_1$ (with hidden mode values $y=0,1$ respectively) dictate whether the next flow rate fluctuation towards zero will result in a pause $P$ or clog $C$. The boxes show example flow rate trajectories from each of these coarse-grained states. 
(b) Distribution of pause count $N$ before a clog. Note the geometrical distribution with $\langle N \rangle = 0.682$.
(c) Distribution of time to clog for trajectories with $N$ pauses. Solid lines are the prediction~\cite[Eq.~(S4)]{SM}. (d) Distribution of time to clog across all trajectories. Lines show an exponential fit (dashed) and the predicted distribution (solid) Eq.~(\ref{doubleExponentialDistribution}), with $\tau_+ = 2.856$~s, $\tau_-=0.580$~s, and $c=0.301$. (e) Distribution of total ejected mass normalized by large grain mass $M/m_L$. Note the exponential distribution and mean $\langle N+1 \rangle m_F = 116.7 m_L$. Model parameters $p$, $\tau_F$, $\tau_P$ and $m_F$ are fit in (b) and \cite{SM}; the predictions in (c)-(e) require no fitting. All data are from the $D=3.86 d_L$ dataset.
}
\end{figure}

For simplicity, we assume the hidden variable $y$ is binary $y=0,1$ with flow-rate-dependent switching rates $\gamma_0(x)$ and $\gamma_1(x)$,
\begin{equation}\label{hiddenVariable}
y: \quad \quad 0 \xrightleftharpoons[\gamma_1(x)]{\gamma_0(x)} 1.
\end{equation}
With our two state variables, $x=y=0$ is the only absorbing state of the dynamics, \textit{i.e.} the clog state. On the other hand, $x\approx 0$, $y=1$ represents a pause, during which the system is susceptible to noise, $\dot x \approx \sqrt{2 \epsilon}\eta(t)$ that eventually perturbs the system back into steady flow. Since spontaneous unclogging is possible even without external vibrations \cite{thomas2016intermittency, janda2009flow, harth2020intermittent}, the additive noise $\epsilon$ has contributions intrinsic to the system and from external vibrations.

The hidden variable switching dynamics must satisfy $\gamma_0(0)=0$ to preserve the absorbing clog state. Expanding to linear order in $x$  and assuming no transitions occur when $x \leq 0$~\footnote{Small negative flow rate is both allowed in our model and occasionally observed in experiment due to particles being pushed backwards (partially) over the mass boundary line, for example due to a mostly static arch shifting to the right or left [see Fig.~\ref{fig:experiment_trajectories_model}(a)]. In both the model and experiment this rate is always small, never comparable to typical positive flow $x_0$. Allowing negative flow and transitions for $x<0$ in the model does not significantly impact our results.}, we have $\gamma_0(x) = (r_0 \cdot x/x_0)H(x)$ and $\gamma_1(x) = (r_{1,0} + r_1 \cdot x/x_0) H(x)$, where $H(x)$ is the Heaviside step function and $x_0$ is the steady flow rate. We assume $r_{1,0}=0$, but our results are robust if $r_{1,0}^{-1}$ is longer than the average pause duration.
Simulated trajectories and flow rate distributions from the stochastic model closely resemble the experimental flows [Fig.~\ref{fig:experiment_trajectories_model}(c)].

To demonstrate the model's ability to quantitatively predict clogging and pausing statistics, we group its internal states by two criteria: flowing versus not flowing, and the value of the hidden mode, $y=0$ versus $y=1$. The result is an analytically solvable four-state representation of our model [Fig.~\ref{fig:flowPauseClogStatistics}(a)] with two indistinguishable flow states $F_0$ and $F_1$ ($x>0$ and $y=0$ and $1$, respectively), a pause state $P$ ($x=0$, $y=1$), and a clog state $C$ ($x=y=0$). Transitions between the four states are Poissonian. Using this representation, we require only four easily measurable parameters to entirely determine the dynamics: the probability of clogging $p$, the average mass ejected during a flow (prior to either a pause or clog) $m_F$, and timescales $\tau_F$ and $\tau_P$, the average flow and pause duration, respectively. In our experiments, flow and pause duration are exponentially distributed, at least for short times, with means $\tau_F=0.77$s and $\tau_P=1.26$s. Flow mass is also exponential, with mean $m_F/m_L =69.4$, where $m_L$ is the large grain mass~\cite{SM}.

During steady flow, the system switches between flow states that lead to clogs ($F_0$) and those that lead to pauses ($F_1$) based on the dynamics of the hidden variable $y$. If, as we assume, the switching of $y$ is fast compared to the flow duration $\tau_F$, the likelihood of a clog is set by $p = r_1/(r_0 + r_1)$. As a result, the number of pauses $N$ before a permanent clog are geometrically distributed,
\begin{equation}\label{geometric}
p(N)  = (1-p)^N p,  \quad \quad \langle N \rangle  = \frac{1-p}{p}.
\end{equation}
This prediction fits our experimental data well [Fig~\ref{fig:flowPauseClogStatistics}(b)], with the single parameter $p=0.595$, so that $\langle N \rangle = 0.682$.

Using the measured clog probability $p$ and timescales $\tau_F$ and $\tau_P$ from experiment, our model accurately predicts---with no additional fitting parameters---a variety of nontrivial statistics of hopper flows. First, it captures the probability distribution of time-to-clog for a flow with exactly $N$ pauses, $p_{T_N}$. This quantity is the sum of $N$ pause and $N+1$ flow timescales, which are each exponentially distributed. We derive the (lengthy) analytical form of $p_{T_N}$ in the Supplemental Material~\cite{SM}. The predicted distribution is in excellent agreement with the data for $N=0,1,2,3$ [Fig.~\ref{fig:flowPauseClogStatistics}(c)]. 

Second, we obtain clogging-time distribution $p_T$ for the entire ensemble of measured flows by averaging $p_{T_N}(t)$ over the geometric distribution of pauses, Eq.~(\ref{geometric}). The result is a double exponential distribution,
\begin{equation}\label{doubleExponentialDistribution}
p_T(t) = \frac{c}{\tau_+} e^{- t/\tau_+} +  \frac{(1-c)}{\tau_-} e^{- t/\tau_-},
\end{equation}
with the derivation and analytic forms of $c$, $\tau_+$, and $\tau_-$ all as functions of $(p,\tau_P,\tau_F)$, included in the Supplementary Material~\cite{SM}. Again without fitting, the prediction Eq.~(\ref{doubleExponentialDistribution}) nicely captures the data, noticeably better than a single exponential [Fig.~\ref{fig:flowPauseClogStatistics}(d)]. We find that $\tau_F$ increases with outlet size $D$, while $\tau_P$ is approximately constant~\cite{SM}. As a result, at large outlet sizes the first exponential ($\tau_+$) dominates \cite{SM}, recovering the established exponentially distributed clogging time \cite{zuriguel2005jamming, to_jamming_2005, janda2008jamming, zuriguel2011silo}.

Third, the model predicts an exponential distribution of ejected masses prior to a clog $p_M$ regardless of outlet regime, consistent with our experiments [Fig.~\ref{fig:flowPauseClogStatistics}(e)] and as well-established in the literature~\cite{zuriguel2005jamming, to_jamming_2005, zuriguel2011silo, thomas2015fraction,koivisto_effect_2017}. The total mass ejected in a trajectory with $N$ pauses is simply the sum of $N+1$ flow masses $m_F$, as no mass is ejected during a pause. Averaging over the geometric number of pauses produces an exponential distribution with predicted mean $\langle M \rangle = m_F \langle N+1 \rangle =116.7 m_L$ [Fig.~\ref{fig:flowPauseClogStatistics}(e)], within $2\%$ of the empirical average mass ejected~\cite{SM}.

We now return to the full Langevin dynamics, Eqs.~(\ref{flowRateDynamics}) and (\ref{hiddenVariable}), for deeper insight into statistics of the flow rate. The flow rate distribution $P_y(x,t)$ for the hidden variable state $y=0,1$ satisfy a coupled Fokker-Planck equation (using the It\^{o} interpretation for multiplicative noise), 
\begin{equation}
\begin{split}
    \frac{\partial P_y}{\partial t} &= -\frac{\partial (f P_y)}{\partial x} + \frac{\partial^2 (g_y P_y)}{\partial x^2} + \mathcal{R}(y\to1-y)  
\end{split}
\end{equation}
where $g_y(x) = g(x)+\epsilon y$ and $\mathcal{R}(y\to1-y) = \frac{x}{x_0}(r_{1-y} P_{1-y} - r_y P_y)H(x)$ describes probability flow between $P_0$ and $P_1$ from dynamics of the hidden variable. The measured flow rate distributions are approximated by the slowest decaying mode of the Fokker-Planck operator $\mathcal{L}_\text{FP}$. We compute this eigenfunction numerically and fit to experiments by tuning the noise $ \Delta \tau_0$, steady flow rate $x_0$, and two additional parameters \cite{SM}.

\begin{figure}[t!]
{\centering \includegraphics[width=0.85\linewidth]{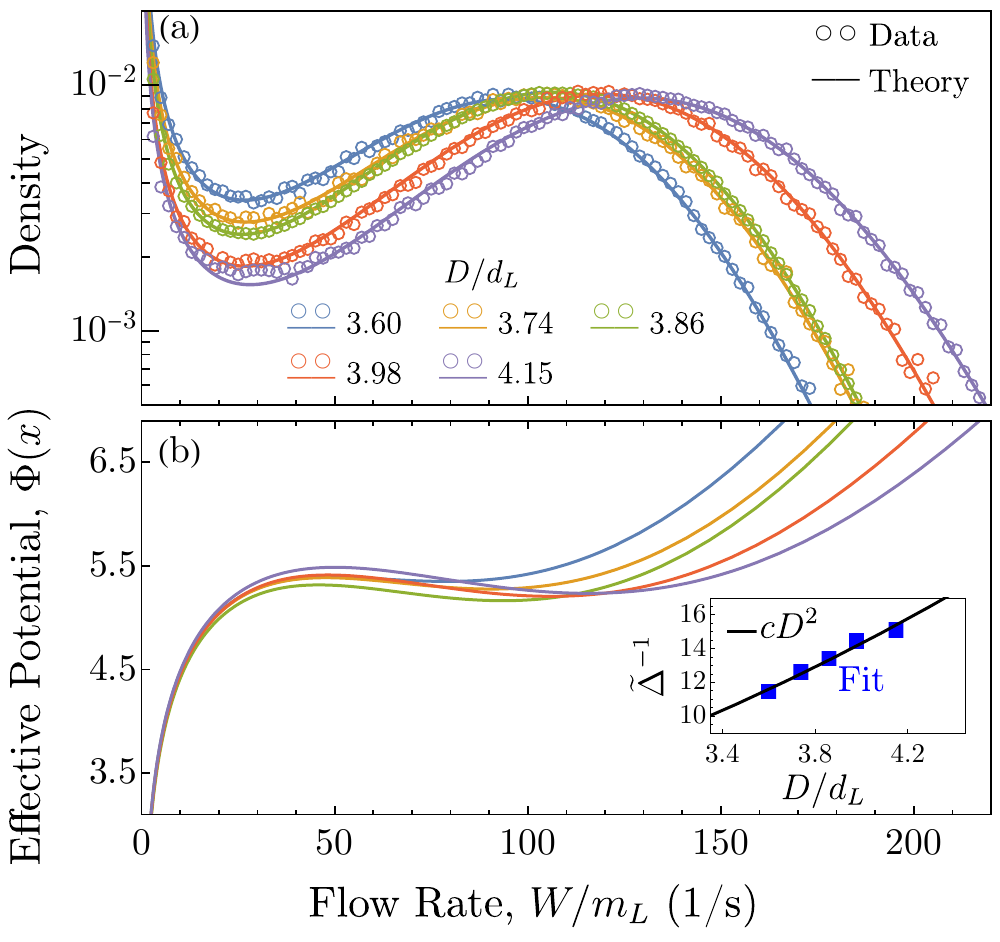}}
\caption{\label{fig:flowRate}
(a) Measured (symbols) and fit (lines) flow rate distributions for various outlet sizes.  
(b) The corresponding effective potentials $\Phi(x)$ and (inset) $D$-dependence of $\widetilde \Delta = \Delta \tau_0/x_0^2$ from fit parameters $x_0$ and $\Delta \tau_0$, which scales quadratically $\widetilde \Delta^{-1} \approx 0.89 (D/d_L)^2$.
}
\end{figure}

Fig.~\ref{fig:flowRate}(a) shows the resulting fits to flow rate distributions for various outlet sizes. Our model explains key features of these small outlet flow rate distributions. Intuitively, smaller outlet sizes promotes slower flows (smaller steady flow rate $x_0$). While noise also shrinks with outlet size, approximately proportionally to $x_0$, relative fluctuations $\widetilde \Delta =\Delta \tau_0/x_0^2$ grow. At small outlets these larger fluctuations and proximity of $x_0$ to the clog state $x=0$ create the non-Gaussian shape in Fig.~\ref{fig:flowRate}(a), while the peak near zero comes from coupling to the hidden mode. Conversely, for large outlets, our model predicts the experimentally observed Gaussian distribution~\cite{janda2009flow, thomas2016intermittency, zhang2023experimental, SM}.

We calculate the corresponding effective potentials for the flow rate dynamics, $\Phi(x) = -\int [f(x)-g'(x)]/g(x) \, dx$ [Fig.~\ref{fig:flowRate}(b)]. The non-monotonic shapes of $\Phi$ illustrate that clogging (and pausing) is a barrier crossing phenomena~\cite{hanggi1990reaction, hathcock2021reaction} with barrier height $\propto \widetilde \Delta^{-1}$; grains flow at a steady rate (the local minimum of $\Phi$) before a large fluctuation pushes the flow rate over the barrier leading to formation of a clog (or pause).

Using first-passage-time analysis, we predict the scaling of the clogging time for large outlets \cite{SM, vanKampen2007stochastic}: $\langle T \rangle \approx \tau_F/p \sim   \tau_0 p^{-1} C \exp(c \widetilde \Delta^{-1})$. Remarkably, this form is compatible with the previously measured stretched-exponential growth of the clogging time~\cite{thomas2015fraction, thomas2016intermittency, to_jamming_2005, janda2008jamming}, $\langle T \rangle \propto  \exp(c D^\alpha)$, where $\alpha$ is the dimension of the hopper ($\alpha = 2$ here). The implied scaling $\widetilde \Delta \propto D^{-2}$ is compatible with our fits [Fig.~\ref{fig:flowRate}(b), inset]; testing this scaling over a broader range of outlets will be an interesting direction for future work.
This result provides further theoretical support for the lack of a critical clogging transition with outlet size  \cite{thomas2015fraction, thomas2016intermittency}. Moreover, it provides a physical interpretation for the effective noise reduction with increasing outlet size. The multiplicative noise term arises due to the discrete nature of the few grains whose configurations determine the flow rate $\widetilde \Delta \propto D^{-2} \sim N_\text{outlet}^{-1}$, which is suppressed as this ensemble grows.

\begin{figure}[t!]
{\centering \includegraphics[width=0.9\linewidth]{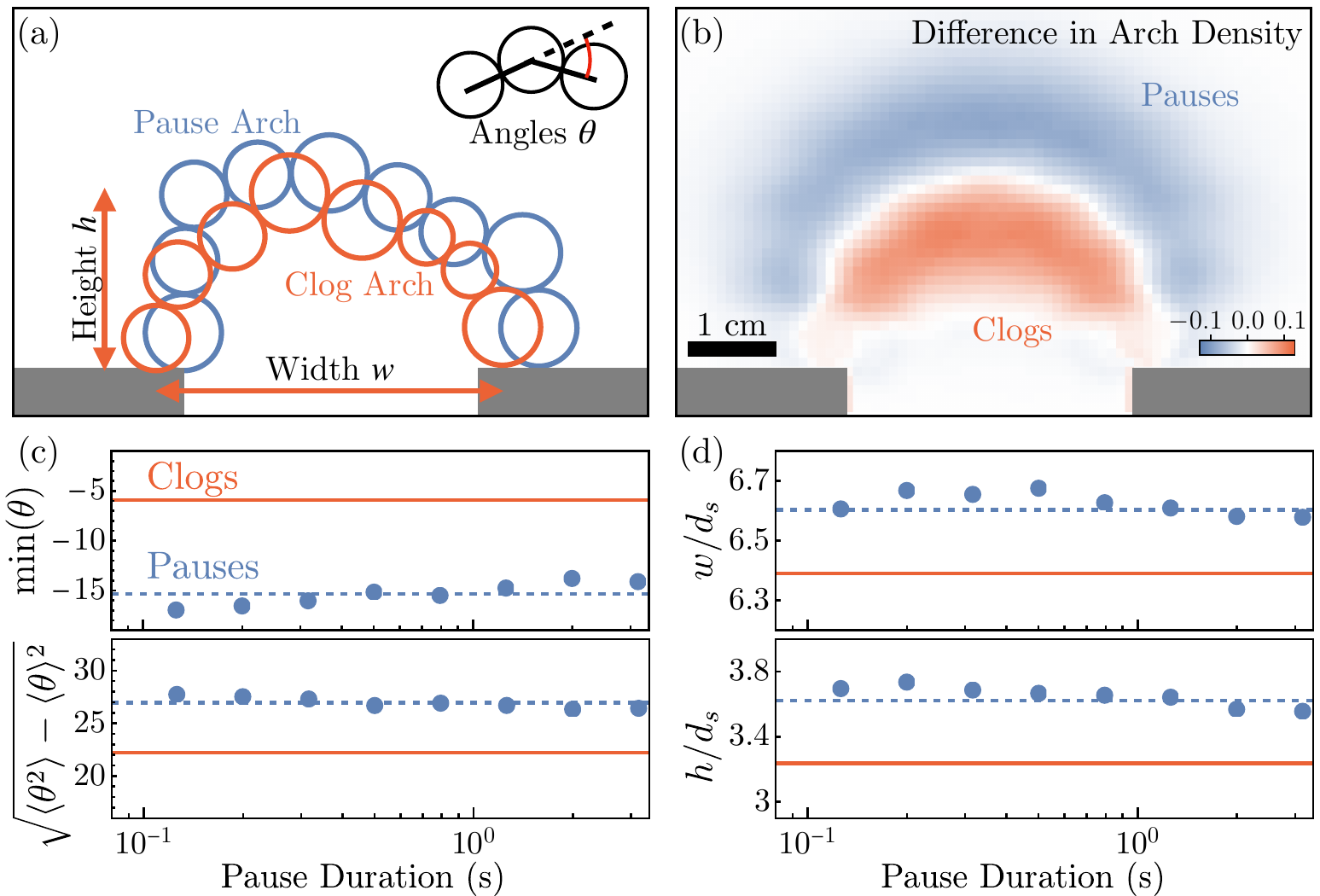}}
\caption{\label{fig:archConfigurations}
\emph{Configurational features of the hidden mode.} (a) Representative pause and clog arches with definitions of the height $h$, width $w$ and angles $\theta$. (b) Difference (clog minus pause) in average positional arch densities. (c) Minimum arch angle and angle standard deviation and (d) arch width and height for pauses and clogs. Each plot shows population average (lines) and pauses binned based on pause duration (points). \revise{For (c) the minimum and standard deviation are computed for each arch before averaging.} Pauses arches are on average wider and taller with more variable curvature.}
\end{figure}

\revise{Finally, we investigate the microscopic origin of the hidden mode. While we do not have direct access to $y$, since it is a coarse-grained representation of degrees of freedom beyond the flow rate, positional snapshots of the grains reveal physical features that characterize the hidden mode. The state of $y$ is uniquely determined in the experiment during moments with no flow ($x=0$) by the long term outcome: whether a clog is temporary ($y=1$) or permanent ($y=0$). For each zero-flow event, we isolate the arch grains [Fig.~\ref{fig:archConfigurations}(a)].} The positional difference between pauses and clogs is immediately apparent from these data. First, by simply averaging reconstructions of the arch grains, we find a clear size difference, with pause arches significantly larger on average than clog arches [Fig.~\ref{fig:archConfigurations}(b)]. We compute various statistics~\cite{zuriguel2020statistical,alborzi2023mixing, guerrero2018slow, alborzi2024faster}: arch width, height, and the variance and minimum of inter-grain angles (details in \cite{SM}), finding further clear contrast [Fig.~\ref{fig:archConfigurations}(c) and (d)]. For these four measures, pauses and clogs differ substantially on average. Further, these results are consistent \textit{regardless of pause length}~\cite{guerrero2018slow}, in a striking confirmation of our model's binary mode separating pauses and clogs. That is, longer and longer pauses do not become statistically equivalent to clogs, they remain distinct. Overall, pause arches are wider, taller, and have a more variable and irregular shape compared to permanent clogs, which have been shown in other hopper systems to be often symmetric~\cite{ahmadi_experimental_2018}. 
\revise{These statistics identify key configurational features that determine the state of the hidden mode in the stochastic model. Our analysis of the model thus demonstrates how these miscrosopic features couple to the macroscopic flow dynamics and noise to determine flow-rate fluctuations, clogging statistics, and intermittency. }

In this Letter, we have established a minimal Langevin model for the dynamics of granular hopper flows. We find that a flow rate fluctuations driven by multiplicative noise and coupled to a configurational mode that determines clog stability explains the statistics of flowing, clogging, and intermittency. A more precise determination of the configurational mode dynamics will be an interesting direction for future work. For example, does increasing $D$ change the switching rates $r_0$, $r_1$ by modifying population of microstates corresponding to clogs and pauses? Is stability always determined at the onset of a clog, or do some pauses become permanent over long timescales via frictional aging and memory effects~\cite{dieterich_timedependent_1972, dillavou_shear_2020}? While it appears configurational features are useful in determining the long-term stability of a clog, these measures only weakly correlate with pause duration [Fig.~\ref{fig:archConfigurations}(c)-(d)]. On this front, previous models of arch breaking and defects~\cite{nicolas2018trap,lozano2012breaking, lozano2015stability} could be incorporated into our model.

It will also be exciting to study the implications of the configurational mode for clog prediction. While the Markovian nature of the dynamics means the current flow rate provides little information for clog prediction, 
monitoring the configurational mode (\emph{e.g.} using the measures identified in Fig.~\ref{fig:archConfigurations}) can provide advance notice of whether an incipient clog will be temporary or permanent~\cite{SM}. Perhaps these insights can be incorporated into machine learning algorithms~\cite{hanlan2024cornerstones} to distinguish the pause- and clog-causing configurations from steady flow.

Our model raises new  questions about which key properties influence dynamics of hopper flows. Here we have only varied the hopper outlet size, but the dependence of clogging on many other parameters, including hopper geometry~\cite{zuriguel2020statistical, zuriguel2005jamming, janda2012flow, thomas2013geometry, thomas2015fraction, janda2008jamming, caitano2021characterization,gago2023effect}, grain properties~\cite{ hafez2021effect,gella2022dual, escudero2022kinematics, zuriguel2011silo, zuriguel2005jamming,thomas2013geometry,zuriguel2020statistical}, and external vibrations~\cite{lozano2012breaking, lozano2015stability,caitano2021characterization} has been explored; does our model capture these behaviors?
Of particular interest is the recent result that increasing vibrations pushes granular flows through a critical transition, beyond which permanent clogging is impossible \cite{caitano2021characterization}. Our model should naturally extend to this case, through vibrational dependence in the additive noise and other parameters. Understanding the dynamics underlying this transition will shed further light on the fundamental requirements for clogging. Beyond granular materials, our phenomenological modeling approach can be tested across a wider range of clogging systems: do clogging dynamics always admit a low-dimensional description? 
\vspace{1em}

\begin{acknowledgments}
This work was partially supported by NSF grants MRSEC/DMR-1720530 and MRSEC/DMR-2309043. D.J.D. and Y.T. thank the Aspen Center for Physics, which is supported by NSF grant PHY-2210452. D.J.D. thanks CCB at the Flatiron Institute, a division of the Simons Foundation, as well as the Isaac Newton Institute for Mathematical Sciences under the program ``New Statistical Physics in Living Matter" (EPSRC grant EP/R014601/1), for support and hospitality while a portion of this research was carried out. S.D. acknowledges support from the University of Pennsylvania School of Arts and Sciences Data Driven Discovery Initiative.

\end{acknowledgments}

\bibliography{references}

\end{document}